\documentclass[11pt,a4paper]{article}
\pdfoutput=1
\usepackage{jheppub}

\title{A deformed conifold with a cosmological constant}

\author{Stanislav Kuperstein}

 \affiliation{Institut de Physique Th\'eorique, Universit\'e Paris Saclay, CEA, CNRS, F-91191 Gif-sur-Yvette, France}

 \emailAdd{stanislav.kuperstein@gmail.com}

\abstract{We find a new regular solution of six-dimensional Einstein's equations with a positive cosmological constant. It has the same isometry group as the (deformed) conifold geometry, and the superpotential approach is used to solve the equations of motion. The space is compact and interpolates between the deformed conifold and the resolved cone with a blown-up four cycle. The deformation/resolution parameters are set by the cosmological constant.}

\begin{document}
\maketitle

\setcounter{footnote}{0}
\setcounter{figure}{0}
\setcounter{equation}{0}

\section{Introduction and Summary}

The warped conifold geometries \cite{Klebanov:1998hh,Klebanov:1999rd,Klebanov:2000nc,Klebanov:2000hb} are very important supergravity backgrounds in the AdS/CFT correspondence. The singular conifold is a Calabi-Yau cone over the five-dimensional
$T^{1,1}$ space with the $S^2 \times S^3$ topology.
Placing D3-branes at the singular tip of the cone 
leads to a smooth $10d$ solution which is dual to an $\mathcal{N}=1$ $4d$ quiver gauge theory \cite{Klebanov:1998hh}. The solution has been since generalized to far more complicated Sasaki-Einstein spaces.
The conic singularity can be smoothed out by blowing up of either the 3-sphere or an even-dimensional cycle ($S^2$ or $S^2 \times S^2$). The former option is called deformation and the latter is known as resolution of the conifold \cite{Candelas:1989js}. The resolved geometries with D3-brane sources describe mesonic branches of the dual gauge theory \cite{Klebanov:2007us,Benvenuti:2005qb}. On the other hand, when D5 branes wrap the collapsing 2-sphere of the deformed conifold, the gauge theory becomes  non-conformal and cascades down to the confining $\mathcal{N}=1$ YM in the deep IR \cite{Klebanov:2000hb}.\footnote{Having ``left-over" D3 sources on the deformed conifold corresponds to the mesonic branch of the gauge theory \cite{Krishnan:2008gx}.}

The Ansatz for the six-dimensional metric including both the deformed and the resolved conifold solutions was written down by Papadopoulos and Tseytlin (PT) in \cite{Papadopoulos:2000gj}. It was also shown there that for the two solutions the Ricci-flatness equations might be solved using the superpotential method. In \cite{Butti:2004pk}  a new one-parameter family of regular IIB supersymmetric solutions was found based on the PT Ansatz. Importantly, the six-dimensional space of these $10d$ backgrounds becomes Ricci-flat only when it approaches the deformed conifold solution. The type IIB backgrounds describe the baryonic branch of the Klebanov-Strassler (KS) cascading gauge theory, along the line of the earlier works \cite{Aharony:2000pp,Gubser:2004qj}. One may also solve the second-order Ricci-flatness equations for the interpolating PT Ansatz, ending up with a non-Calabi-Yau metric \cite{Dymarsky:2011ve}. The corresponding $10d$ solution describes the KS gauge theory perturbed by certain combinations of relevant single trace and marginal double trace operators.

The goal of this paper is to find \emph{Einstein-flat} solutions based on the PT Ansatz. To be more precise, we focus on the deformed conifold part of the Ansatz, which apart from the $SU(2) \times SU(2)$ preserves an additional $\mathbf{Z}_2$ symmetry. Using the superpotential method we find a new solution parametrized by the (positive) cosmological constant. The $6d$ space is everywhere \emph{regular} and, as expected, compact. For one limiting value of the (former) radial coordinate the space asymptotes to the deformed conifold solution, while in the other limit one finds a regular resolution with the blown up $S^2 \times S^2$ four-cycle. Remarkably, the regularity at either end requires no orbifolding of the $5d$ base. 
The sizes of the 3-sphere and the 4-cycle are related and both are determined by the cosmological constant $\Lambda$. The metric has a singular limit, where (at least) one corner of the space has a conic singularity. The singular solution 
preserves the $U(1)_\psi$ symmetry associated with the Reeb angle of the $T^{1,1}$ base.
Although our space is compact, we will still refer to the regions with the deformed and the resolved spaces as the IR and the UV respectively. The reason for these notations will be clarified later in the paper.

The Ansatz we consider involves three independent functions of the radial coordinate, and in general requires solving a set of three second-order coupled non-linear ordinary differential equations (ODEs), similar to \cite{Dymarsky:2011ve}. However, the superpotential we found greatly simplifies the task, since it leads to \emph{first}-order ODEs. Out of the three equations, one can be solved analytically and the other two combine into a single second-order equation that can be treated numerically. This allows us to find a relation between the ``UV" resolution and the ``IR" deformation parameters. Importantly, the superpotential approach is futile for the full PT Ansatz, and this is the main reason we imposed the $\mathbf{Z}_2$ symmetry. As a result, we have to exclude the 2-cycle resolution from the discussion, since it does not fit into the $\mathbf{Z}_2$-symmetric Ansatz.

There are three  primary motivations for this work. \emph{First}, it provides a natural extension of the well-known old results. The so-called \emph{Eguchi-Hanson-de Sitter} space was found more than 30 years ago in \cite{Gibbons:1979xn,Pedersen}. It is the compact version of the Eguchi-Hanson (EH) geometry \cite{Eguchi:1978xp}, which solves Einstein equations with a positive cosmological constant. At both ``corners" of the space the $\mathbb{C}^2/\mathbf{Z}_2$ singularity is resolved by blown-up two-spheres of the same size. This size is, in turn, fixed by the cosmological constant. This is very similar in spirit to the results of this paper, though we find cycles of different dimensions in the IR and the UV. This should be of no surprise, since the 2-cycle resolution/deformation is the only option in four dimensions.\footnote{Recall that the 2-cycle resolution is not included in our Ansatz for reasons explained above.}

\emph{Second}, consistent Kaluza-Klein reductions of type IIB supergravity on the compact six-dimensional space may lead to interesting four-dimensional gauged supergravity theories. The truncation will be probably easier for the singular version of the solution due to the preserved $U(1)_\psi$ symmetry factor. The reductions (if exist) will share many features with the one constructed in
\cite{Cassani:2010na, Bena:2010pr}.\footnote{See also \cite{Cassani:2012wc} for a possible interpretation of the new solutions as curved domain walls in the truncated supergravity theories.}

\emph{Third}, our solution can be used to build a new \emph{KS-like} background  in type IIB supergravity with  $\textrm{AdS}_4$ and our compact $6d$ geometry replacing $\textrm{Mink}_4$ and the non-compact deformed conifold respectively.
Solutions with 3-form imaginary self-dual (ISD) fluxes\footnote{We follow the conventions of \cite{Bena:2014jaa}, with $D3$- and $\overline{D3}$-branes being mutually supersymmetric with ISD and IASD fluxes respectively.} on compact Einstein-flat transversal spaces have recently attracted a great deal of attention (see \cite{Gautason:2015ola} for the most recent developments). The 5-form tadpole cancellation on the compact $6d$ space necessitates the introduction of either anti-D3 brane sources or orientifolds.  Supergravity solutions with anti-branes placed in backgrounds that contain opposite charges dissolved in the fluxes are known to have certain singularities (see \cite{Bena:2014jaa} for the extended list of references). For example, anti-D3 branes \emph{smeared} over the tip of the warped deformed conifold induce a 3-form flux singularity, as was proven in \cite{Bena:2012bk}. It was furthermore argued in \cite{Bena:2012vz} that this singularity cannot be cured by Polchinksi-Strassler polarization \cite{Polchinski:2000uf} of D5-branes warping the shrinking 2-sphere. The situation may, however, change once the deformed conifold is made compact. As a toy model capturing some of the physics, one may consider the anti-D6 singularity in massive type IIA supergravity \cite{Blaback:2011pn}. This is the (three times) T-dual of the anti-D3's  we discussed above, but now smeared over a 3-torus rather than over the $S^3$ at tip of the KS geometry. According to \cite{Bena:2012tx}, for flat Minkowskian world-volume, the singularity cannot be resolved by D8-brane polarization independently of the parameters of the fully backreacted anti-D6 solution.
When the D6's have $AdS_7$ world-volume, however, the polarization fate depends on the values of the cosmological constant and other parameters of the fully backreacted anti-D6 solution.\footnote{The $AdS_7$ solution was later found in \cite{Apruzzi:2013yva}.} Hence, it will be exciting to see whether our ``compact deformed conifold" space leads to a 3-form flux singularity that can be smoothed out by the 5-brane polarization. To answer this question one will have to understand first whether the cosmological constant is a free parameter in the fully backreacted solution or it is rather determined by the fluxes, as it happens for completely smeared sources  \cite{Blaback:2010sj}. The method presented in  \cite{Gautason:2013zw} might appear useful to answer this question without constructing the full solution.

The fact that our solution follows from a superpotential strongly suggests that once embedded in the type IIB supergravity background it will preserve some amount of supersymmetry which will be further broken by the anti-branes. We leave this interesting question for the future research.

The paper is organized as follows. In Section \ref{sec:Ansatz} we present the metric Ansatz, the one-dimensional effective action, the superpotential equation solution and the corresponding first-order equations of motion. We also briefly review the known non-compact solution with zero cosmological constant.  In Section \ref{sec:NewSolutions} we write down the new compact solutions, both singular and regular. In the Appendix we give the 1-forms definitions and summarize the relation to the conventions of \cite{Papadopoulos:2000gj}.

\section{The Ansatz for the metric and the equations of motion}
\label{sec:Ansatz}

Our Ansatz for the six-dimensional metric has the same isometries as the deformed conifold space of \cite{Candelas:1989js}, which in turn is a particular example of the Papadopoulos-Tseytlin (PT) metric Ansatz \cite{Papadopoulos:2000gj}:
\begin{equation}
\label{eq:metric6d}
\textrm{d} s_{6}^2 = \frac{2}{3} e^{-2 z + w} \left( \textrm{d} \tau^2 + g_5^2 \right)
       +  e^{z} \Big( e^y \left(  g_1^2 + g_2^2 \right)  +  e^{-y} \left(  g_3^2 + g_4^2 \right) \Big) \, .
\end{equation}
Here the functions $z(\tau)$, $w(\tau)$ and $y(\tau)$
depend only on the radial coordinate $\tau$, and the definitions of the angular one forms $g_i$ are given in Appendix \ref{sec:1-forms}. 
Apart from the $SU(2) \times SU(2)$ isometry, for $y=0$ the metric enjoys an additional $U(1)_\psi$ symmetry associated with the Reeb angle $\psi$.  For a non-zero $y(\tau)$ the $U(1)_\psi$ is broken down to  $\mathbf{Z}_2$.
An extra $\mathbf{Z}_2$ symmetry preserved by (\ref{eq:metric6d}) acts on the angles as $g_{1,2} \to - g_{1,2}$ with the other three 1-forms being invariant. In terms of the angles in (\ref{eq:1-forms}) it is merely $\left( \theta_1, \phi_1 \right) \leftrightarrow \left( \theta_2, \phi_2 \right)$. This symmetry reduces by one the number of functions in the most general PT Ansatz.\footnote{The 2-cycle resolved conifold with the blown-up $S^2$ breaks the latter $\mathbf{Z}_2$, but preserves the $U(1)_\psi$ \cite{Candelas:1989js}.}  We relegated to Appendix \ref{sec:PT} the relations between our functions and those of \cite{Papadopoulos:2000gj}.

The most general \emph{regular} \emph{Ricci-flat} solutions of the form (\ref{eq:metric6d}) are the deformed conifold metric \cite{Candelas:1989js} and the 4-cycle resolution of the $T^{1,1}/\mathbf{Z}_2$ singularity, and we will review both solutions in the next section. 
We are, however, interested in an \emph{Einstein-flat} solution. To obtain the one-dimensional effective action for the three scalar functions one has to plug (\ref{eq:metric6d}) into the Einstein-Hilbert action and integrate over the five angles. The output is:
\begin{eqnarray}
\label{eq:Lagrangian}
\qquad  - \frac{1}{2} G_{a b}(\phi) {\phi^a}^\prime  {\phi^b}^\prime  - V(\phi) =   &&
\\
= -\frac{3}{4} e^{2 z} \left( {z^\prime }^2 - 2 z^\prime w^\prime  + {y^\prime}^2 \right) &-& \left( \frac{1}{3} e^{-4 z + 2 w} - 2 e^{-z + w} \cosh y + \frac{3}{4} e^{2 z} \sinh^2 y + \frac{ e^{w}}{R^2}\right) 
 \nonumber \, ,
\end{eqnarray}
where the last terms comes from the cosmological term $\sqrt{g_6} \Lambda$ in the action with:
\begin{equation}
\label{eq:CS}
\Lambda = \frac{6}{R^2} \, ,
\end{equation}
and the remaining terms can be found in \cite{Papadopoulos:2000gj}. Let us stress again that in our quest for a \emph{compact} solution we need a \emph{positive} cosmological constant as it appears in (\ref{eq:CS}). By the end of the last section we will comment on the non-compact solution for the negative $\Lambda$.
 
Surprisingly the inclusion of the new term still allows for a simple solution of the superpotential equation for (\ref{eq:Lagrangian}):\footnote{We follow the following conventions for the superpotential equation and the first-order equations of motion:
\begin{equation}
V = \frac{1}{8} G^{a b} \frac{\partial W}{\partial \phi^a}\frac{\partial W}{\partial \phi^b}
\, , \qquad
{\phi^a}^\prime = \frac{1}{2}  G^{a b} \frac{\partial W}{\partial \phi^b} \, .
\end{equation}}
\begin{equation}
\label{eq:W}
W \left( z, w, y \right) = -2 e^{-z + w} -  3 e^{2 z} \cosh y + \frac{ e^{3 z}}{R^2} \, .
\end{equation}
Let us stress that (\ref{eq:W}) is \emph{not} the most \emph{general} solution of the superpotential equation. Typically  (\ref{eq:W}) should be a special case of a solution depending on (maximum) two free parameters, but we were not able to find it.

The superpotential (\ref{eq:W}) leads to the following equations of motion:
\begin{equation}
\label{eq:EOMs-1}
z^\prime = \frac{2}{3} e^{-3 z + w} \, \qquad
w^\prime  = 2 \cosh y - \frac{e^z}{R^2} \, \qquad
y^\prime  = - \sinh y \, .
\end{equation}
Remarkably, the equation for $y(\tau)$ has no $R$ in it and so has exactly the same solutions as for the Ricci-flat metric \cite{Candelas:1989js,Papadopoulos:2000gj}:
\begin{equation}
\label{eq:y-sol}
y = 0 \qquad \textrm{and} \qquad   e^y = \tanh \frac{\tau}{2} \, .
\end{equation}
The first solution preserves the $U(1)_\psi$, while the second one breaks it down to $\mathbf{Z}_2$  (see the comment below (\ref{eq:metric6d})). In what follows we will consider both options for $\Lambda=0$ as well as for $\Lambda>0$. We will see that the $U(1)_\psi$-breaking choice of $y(\tau)$ yields a regular (deformed) solution both for the non-compact and the compact solutions.

The first two equations in (\ref{eq:EOMs-1}) can be recast in the following form:
\begin{eqnarray}
\label{eq:MainEq}
\left( e^{3 z}\right)^{\prime \prime} - 2 \cosh y \left( e^{3 z}\right)^{\prime} + \frac{3}{4 R^2} \left( e^{4 z}\right)^{\prime} &=& 0 
\\ 
e^{w}  = \frac{1}{2} \left( e^{3 z} \right)^\prime
\, . \qquad \qquad &&
\end{eqnarray}
We see that the superpotential method leads to a single second-order equation of motion. Solving (\ref{eq:MainEq})
for $z(\tau)$ determines $w(\tau)$ directly from the remaining equation, while the third function solution is given in (\ref{eq:y-sol}).

Before proceeding to the new solution with a non-zero cosmological constant, let us briefly review the known Ricci-flat solutions arising from (\ref{eq:MainEq}) for $R \to \infty$. 
For the $U(1)_\psi$-symmetric choice, $y(\tau)=0$, one easily finds that:
\begin{equation}
\label{eq:Z2noncomp-sol}
e^z = \frac{r^2}{6} \, , \qquad 
e^{w} = \frac{1}{216} \left( r^6 - r_0^6 \right) \, \qquad
\textrm{for} \qquad
e^{2 \tau} = \left( \frac{r}{r_0} \right)^6 - 1 \, .
\end{equation}
The 6-dimensional metric is then:
\begin{equation}
\label{eq:Z2noncomp-metric}
\textrm{d} s_{6}^2 =  \left( 1 - \left( \frac{r_0}{r} \right)^6 \right)^{-1}\textrm{d} r^2 + \frac{r^2}{9} \left( 1 - \left( \frac{r_0}{r} \right)^6 \right) g_5^2 + \frac{r^2}{6} \left(  g_1^2 + g_2^2 +  g_3^2 + g_4^2 \right) \, .
\end{equation}
For $r_0=0$ this reduces to the singular conifold metric, while for a non-zero $r_0$ it describes a geometry with a blown up 4-cycle, which is just the product of the spheres, $S^2 \times S^2$, spanned by $\left( \theta_1, \phi_1 \right)$ and $\left( \theta_2, \phi_2 \right)$. Zooming near $r=r_0$ one finds that in order to avoid a singularity, the Reeb angle $\psi$ (see the definition of $g_5$ in (\ref{eq:1-forms})) has to be $2 \pi$-periodic. Since for the singular conifold the period is $4 \pi$, the five-dimensional base is then $T^{1,1}/\mathbf{Z}_2$. Similar solutions exist also for the $Y^{p,q}$ and $L^{a,b,c}$ Sasaki-Einstein spaces. In all these examples the blowing up of the 4-cycle resolves the conic singularity at the tip provided the Reeb angle has the right periodicity.

\begin{figure}[t]
\centering
\includegraphics[trim = 0mm 130mm 0mm 30mm, scale=0.4]{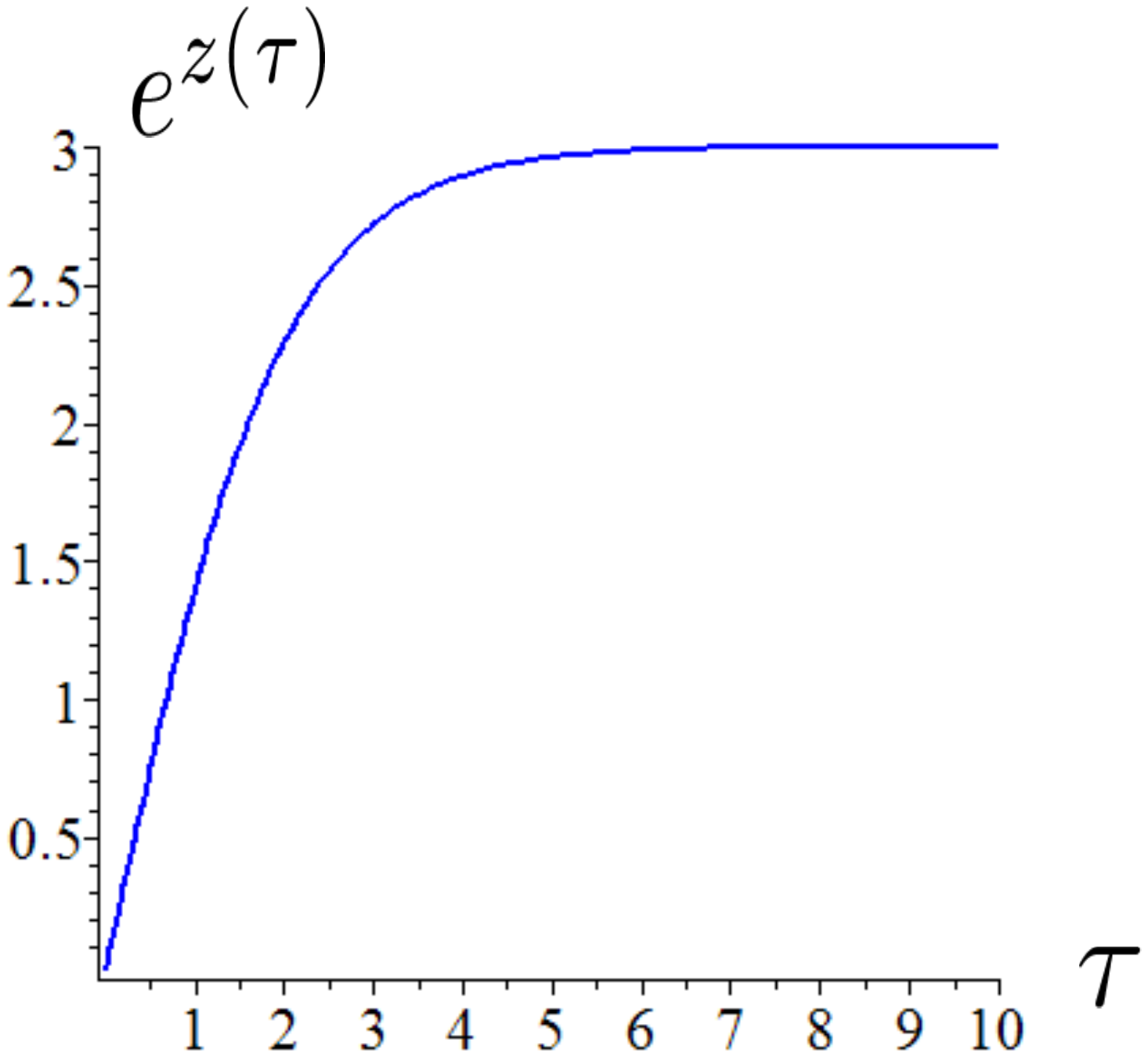}
\caption{The graph shows the numerical solution of (\ref{eq:MainEq}) for $R=1$. For small and large $\tau$ the solution matches (\ref{eq:IR}) and  (\ref{eq:UV}) respectively. The results for the constants  $C_\textrm{\tiny{IR}}$  and $C_\textrm{\tiny{UV}}$  are given below (\ref{eq:UV}).}
\label{fig:Graph}
\end{figure}

For the $U(1)_\psi$-breaking choice in (\ref{eq:y-sol}) the only regular solution is the deformed conifold \cite{Candelas:1989js}:
\begin{equation}
\label{eq:KS-sol}
e^{z} = 2^{-\frac{11}{6}} 3^\frac{1}{4} \epsilon^{4/3} \left( \sinh(2 \tau) - 2 \tau \right)^{1/3} \, , \qquad 
e^{w} = 2^{-\frac{9}{2}} 3^\frac{3}{4} \epsilon^4 \sinh^2 \tau  \, .
\end{equation}
Out of the two integration constants the first one is the deformation parameter $\epsilon$, which measures the $S^3$ size at the tip, and the second constant has to be fixed to avoid a singularity at $\tau=0$.

Before closing up this section it is worth noticing here that for  $\Lambda=0$ (or equivalently infinite $R$) the EOMs (\ref{eq:EOMs-1}) are invariant under $\left( z, w \right) \to \left( z + 2 \lambda, w + 6 \lambda \right)$. For the singular conifold solution this rescaling can be absorbed in  the radial coordinate redefinition, while for the two non-conic solutions it changes the physical size of the corresponding blown-up cycles. We will return to this issue at the end of the next section.

\section{New compact solutions, singular and regular}
\label{sec:NewSolutions}

In this section we will consider solutions with finite $R$. Until the very end of this section we will focus on compact solutions corresponding to the positive cosmological 
constant (see (\ref{eq:CS}) and below). The negative choice of $\Lambda$ produces non-compact geometries as we will discuss soon.

For $y=0$ (unbroken $U(1)_\psi$) the $z(\tau)$ equation (\ref{eq:MainEq}) can be solved analytically:\footnote{The $z(\tau)$ equation (\ref{eq:MainEq}) reduces to $\left( e^{3 z}\right)^{\prime} - 2 e^{3 z}  + \frac{3}{R^2} e^{4 z} = \textrm{const} $, but the constant has to be set to zero in order to avoid a singularity at $\tau=0$. This leaves only one integration constant, $\tau_0$.} 
\begin{equation}
\label{eq:Z2comp-sol}
e^{z} = 8 R^2 \frac{ e^{\frac{2}{3} (\tau-\tau_0)}}{ 1 + 3 e^{\frac{2}{3}(\tau-\tau_0)} } \, , \qquad 
e^{w} = 512 R^6 \frac{e^{2(\tau-\tau_0)}}{\left( 1 + 3 e^{\frac{2}{3}(\tau-\tau_0)}  \right)^4} \, .
\end{equation}
Upon the redefinition
\begin{equation}
e^{\frac{1}{3} \left( \tau -\tau_0 \right)} = \frac{1}{\sqrt{3}} \tan \left( \frac{\alpha}{4} \right) \, ,
\end{equation}
the metric takes the following form:
\begin{equation}
\label{eq:Z2comp-metric}
\textrm{d} s_{6}^2 = R^2 \cdot \left[ \textrm{d} \alpha^2 + \frac{4}{9} \sin^2 \left( \frac{\alpha}{2} \right) \cdot g_5^2 + \frac{8}{3} \sin^2 \left( \frac{\alpha}{4} \right) \cdot  \left(  g_1^2 + g_2^2 +  g_3^2 + g_4^2 \right) \right] 
\end{equation}
with $\alpha \in \left[ 0, 2 \pi\right]$. Near $\alpha=0$ the $T^{1,1}$ part of the metric shrinks and the space looks like the singular conifold geometry. On the other hand, the four cycle has a finite size at $\alpha=\pi$. Expanding near this point we find that the $g_5$ part has no conical deficit provided $\psi$ is $\frac{2}{3} \pi$-periodic, implying that we have to quotient $T^{1,1}$ by $\mathbf{Z}_6$ in order to end up with a regular space at $\alpha=\pi$. The geometry will, however, be still singular at $\alpha=0$.

We will now study the main subject of this paper: the  $U(1)_\psi$-breaking solution with ${e^y=\tanh \dfrac{\tau}{2}}$ and a non-zero cosmological constant. The regular solution for small $\tau$ is:
\begin{equation}
\label{eq:IR}
e^z = C_\textrm{\tiny{IR}} \cdot  \tau + \frac{C_\textrm{\tiny{IR}} }{30} \left( 2 - 3 \frac{C_\textrm{\tiny{IR}}}{R^2} \right) \cdot  \tau^3 + \mathcal{O} \left( \tau^5 \right) \, .
\end{equation}
At leading order it coincides with the deformed conifold solution (\ref{eq:KS-sol}) with $C_\textrm{\tiny{IR}}  \sim \epsilon^{4/3}$. The constant  $C_\textrm{\tiny{IR}}$ is a free IR parameter that has to be properly adjusted by the large-$\tau$ boundary conditions,\footnote{We will refer to the small and the large $\tau$ regions as the IR and the UV even though the space is now compact. The main reason for that is (\ref{eq:y-sol}), which is the same as for the non-compact solution, where large $\tau$ corresponds to the UV region.} since for a generic $C_\textrm{\tiny{IR}}$ the solution will be singular for large $\tau$. Note that according to (\ref{eq:EOMs-1}) $e^z$ and $e^{w}$ are both monotonic increasing functions. This means that the 4-cycle spanned by $g_{1,2,3,4}$ acquires a non-zero size for large $\tau$. At the same time, the equations of motion imply that the $\psi$ 1-cycle shrinks there. In other words, for large $\tau$ the geometry is that of (\ref{eq:Z2noncomp-metric}) with a non-zero $r_0$. Since the periodicity of $\psi$ is already fixed in $\tau=0$ to be $4 \pi$, the space is regular if and only if the $g_5$ part of the metric (\ref{eq:metric6d}) looks asymptotically as $e^{-\tau} \left( \textrm{d} \tau^2 + g_5^2 \right)$.\footnote{Upon the definition $r=e^{\frac{\tau}{2}}$ one gets $4 \textrm{d} r^2 + r^2 g_s^2$ implying that $\psi$ is indeed $4 \pi$-periodic.}
This is, in turn, possible only if $\left( e^z \right)^\prime$ behaves at infinity as $e^{-\tau}$. Such a solution indeed exists and its asymptotic expansion for large $\tau$ is:
\begin{equation}
\label{eq:UV}
e^z = 3 R^2 \left( 1 + C_\textrm{\tiny{UV}} e^{-\tau} +  \frac{1}{2} C_\textrm{\tiny{UV}}^2 e^{-2 \tau}\right) + \mathcal{O} \left( e^{- 3 \tau}\right) \, .
\end{equation}
We finally conclude that the IR integration constant  $C_\textrm{\tiny{IR}}$ has to be chosen such that in the UV the function $e^{z(\tau)}$ approaches the value  $3 R^2$. This will guaranty that the $S^1_\psi$ shrinks there smoothly.

As the equation (\ref{eq:MainEq}) does not allow for an analytic solution for the $U(1)_\psi$-breaking choice of $y$, see (\ref{eq:y-sol}), we have to use numerics to solve this equation. The output for the $e^{z(\tau)}$ function is shown on Figure \ref{fig:Graph}. For $R=1$ the right ``UV" solution is obtained for $C_\textrm{\tiny{IR}}=1.5(0)$. Moreover, matching the numerical solution to the subleading terms of (\ref{eq:UV}) we find that $C_\textrm{\tiny{UV}}=1.9(6)$.

\begin{figure}[t]
\centering
\includegraphics[trim = 0mm 145mm 0mm 20mm, scale=0.45]{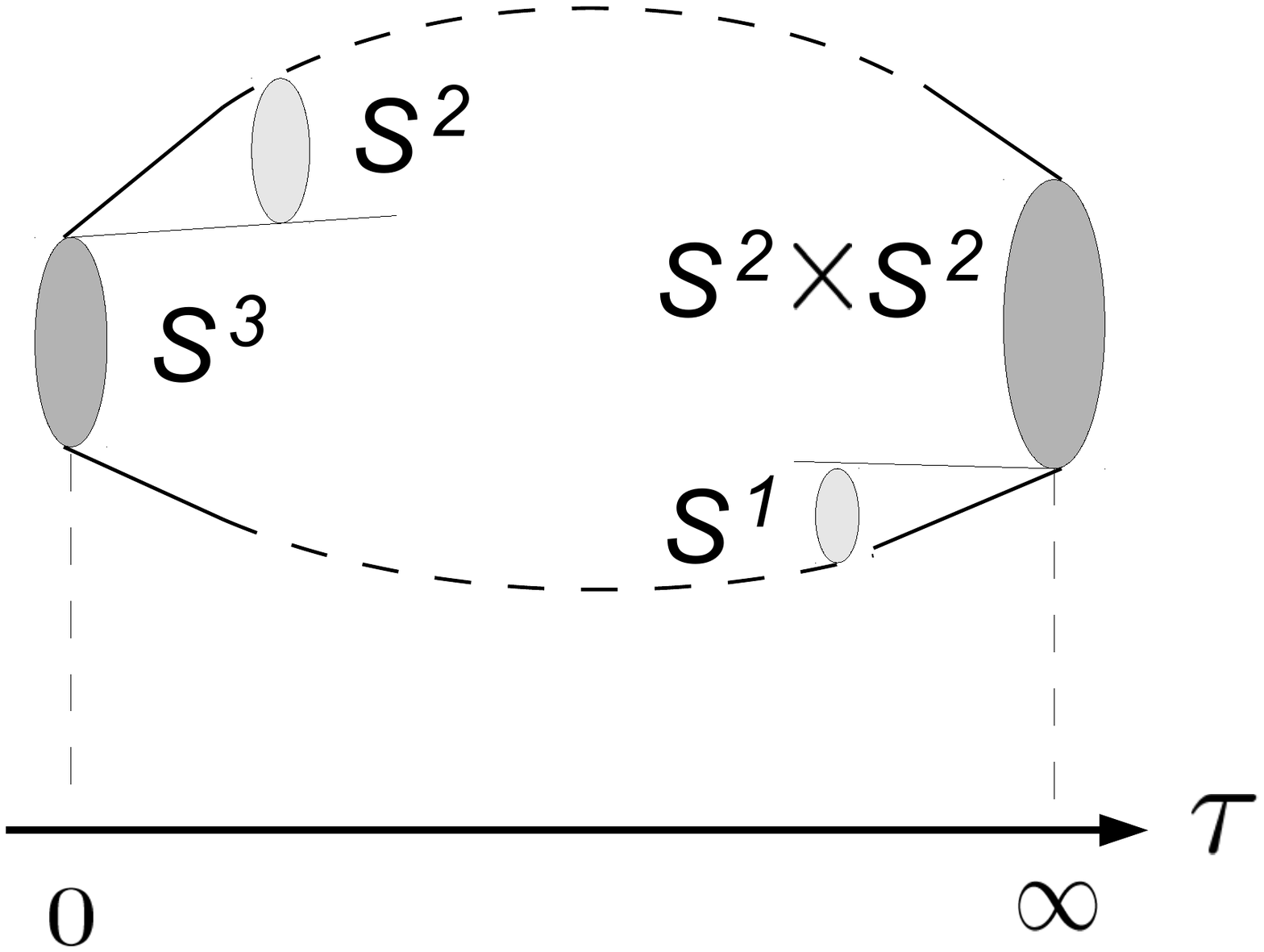}
\caption{The solution on Figure \ref{fig:Graph} describes a compact \emph{regular} space. The ``radial" coordinate $\tau$ is positive, $\tau \in \big[ 0, \infty \big)$. Near $\tau=0$ the 3-sphere is blown up, while the 2-sphere smoothly shrinks. For large $\tau$, on the other hand, the geometry caps off with the finite size $S^2 \times S^2$ four cycle  and the regularly shrinking Reeb vector $S^1$. }.
\label{fig:Shape}
\end{figure}

To summarize, we have found a new regular solution of Einstein's equations with a non-zero cosmological constant having  the isometries of the deformed conifold. 
The space is schematically presented on Figure \ref{fig:Shape}. At the minimal value of the (former) radial coordinate the geometry looks asymptotically like the deformed conifold with the blown 3-sphere and (regularly) shrinking 2-sphere. On the other side, the Reeb vector cycle shrinks instead and the 4-cycle, $S^2 \times S^2$, has a finite size. The size of the blown up $S^3$ at one corner of the space and the 4-cycle size at the other one are related, and both are uniquely fixed by the cosmological constant, $\Lambda = 6 R^{-2}$.  

In fact, the constant $C_\textrm{\tiny{IR}}$, the $\tau=0$ deformation parameter, behaves as $C_\textrm{\tiny{IR}} \sim R^2$. This immediately follows from (\ref{eq:IR}) either by using the dimensional analysis of (\ref{eq:metric6d}) or by noting that for finite $R$ the scaling symmetry mentioned at the end of the previous section modifies to:
\begin{equation}
\left( z, w, R \right) \to \left( z + 2 \lambda, w + 6 \lambda, e^\lambda R \right) \, .
\end{equation}
Together with the numerical result for $R=1$ it implies that
\begin{equation}
C_\textrm{\tiny{IR}}=1.5(0) \cdot R^2 \, .
\end{equation}
Similar analysis reveals also that  $C_\textrm{\tiny{UV}}$
does not scale with $R$ or, in other words, the $R=1$ result, $C_\textrm{\tiny{UV}}=1.9(6)$, holds actually for any $R$.

It is worth to emphasize again that (\ref{eq:W}) is supposedly not the most general solution of the superpotential equation. It is reasonable to believe that there exists a solution for which the sizes of the blown-up cycles depend on additional \emph{free} parameter(s) and not only on $R$. Playing with these parameters it should be (presumably) possible to obtain the singular metric (\ref{eq:Z2comp-metric}) as a special limit of the regular solution. It remains to be seen whether such a general solution does exist, and if yes, whether it follows from a certain superpotential.

Let us finally comment on the $\Lambda<0$ possibility. The $z$ and $y$ equations are not modified, while for the $w$-equation one has to change the sign in front of the second ($R$-dependent) term in (\ref{eq:EOMs-1}). For $y=0$ the solution is the one in (\ref{eq:Z2comp-metric}) with $\sin$'s being replaced by $\sinh$'s and $\alpha$ ranging from $0$ to infinity. For small $\alpha$ the space then has the same conic singularity as for the $\Lambda>0$ choice, but for large $\alpha$ one finds an exponentially divergent geometry. For the broken $U(1)_\psi$ case (non-zero $y(\tau)$) the non-compact solution interpolates in turn between the divergent geometry and the deformed conifold region.

\acknowledgments{We would like to thank Mariana Gra\~na and Thomas Van Riet for stimulating discussions. We are also grateful to Johan Bl{\aa}b{\"a}ck for explaining the results of \cite{Blaback:2010sj} and other related references.
It is a pleasure to thank Iosif Bena for reading through the manuscript and making numerous helpful suggestions and corrections.
This work was supported in part by the ERC Starting
Grant 240210 String-QCD-BH and by a grant from the Foundational Questions Institute (FQXi) Fund, a donor advised fund of the Silicon Valley Community Foundation on the basis of proposal FQXi-RFP3-1321 (this
grant was administered by Theiss Research).}

\appendix

\section{Angular one-forms}
\label{sec:1-forms}

In this appendix we summarize the definitions of the metric 1-forms in terms of the angular coordinates $\theta_{1,2}$, $\phi_{1,2}$ and $\psi$.
\begin{equation}
\label{eq:g-forms}
g_1 = \frac{e_2 - \epsilon_2}{\sqrt{2}} \, , \quad
g_2 = \frac{e_1 - \epsilon_1}{\sqrt{2}} \, , \quad
g_3 = \frac{e_2 + \epsilon_2}{\sqrt{2}} \, , \quad
g_4 = \frac{e_1 + \epsilon_1}{\sqrt{2}} \, , \quad
g_5 = \tilde{\epsilon}_3   \, ,
\end{equation}
where:
\begin{eqnarray}
\label{eq:1-forms}
e_1 = \textrm{d} \theta_1 &&
e_2 =  - \sin \theta_1 \textrm{d} \phi_1
\nonumber \\
\epsilon_1 =  \sin \psi \sin \theta_2 \textrm{d} \phi_2 + \cos \psi \textrm{d} \theta_2 &&
\epsilon_2 =  \cos \psi \sin \theta_2 \textrm{d} \phi_2 - \sin \psi \textrm{d} \theta_2
\nonumber \\
\tilde{\epsilon}_3 = \epsilon_3 + \cos \theta_2 \textrm{d} \phi_2  &=& \left( \textrm{d} \psi + \cos \theta_1 \textrm{d} \phi_1 \right) + \cos \theta_2 \textrm{d} \phi_2 \, .
\end{eqnarray}

\section{Relation to other conventions  in the literature}
\label{sec:PT}

Here we present the connection between the functions of the Ansatz (\ref{eq:metric6d}) and the metric functions used in \cite{Papadopoulos:2000gj}:
\begin{eqnarray}
\label{eq:PTtoHERE}
\left( e^x , e^{6 p}, e^{6 A} \right)_{\rm PT}
    &=& \left( 2^{-\frac{1}{2}} 3^{-\frac{1}{4}} e^{z} , 3^{\frac{3}{2}} e^{z - w}, 2^{-\frac{3}{2}} 3^{-\frac{9}{4}} e^{2 z + w} \right)_{\rm here} \, , \nonumber \\
e^g &=& \frac{1}{\cosh (y)}    \, ,  \quad
a   = \tanh (y)    \, .
\end{eqnarray}
Notice that the relation between $a$ and $g$ is required by the $\mathbf{Z}_2$ symmetry we discussed below (\ref{eq:metric6d}).

\bibliographystyle{utphys}
\bibliography{conifold}

\end{document}